\documentclass[11pt,twoside]{article}
\usepackage{cozumel2005}
\usepackage{graphicx}
\usepackage{epsf}
\usepackage{psfig}
\usepackage{lscape}
\usepackage{rotating}

\begin{document}
\title{Detailed Chemical Abundances of Extragalactic Globular Clusters} 
\author{R.~A.~Bernstein}
\affil{Astronomy Dept, 
University of Michigan, 500 Church St., Ann Arbor, MI, 48109}
\author{ A.~McWilliam}
\affil{Carnegie Observatories, 813 Santa Barbara St., Pasadena, CA, 91101}

\begin{abstract}

We outline a method to measure the detailed chemical composition of
extragalactic (unresolved) globular clusters (GCs) from echelle
spectra of their integrated light.  Our goal is to use this method to
measure abundance patterns of GCs in distant spiral and elliptical
galaxies to constrain their formation histories.  To develop this
technique we have obtained a ``training set'' of integrated--light
spectra of resolved GCs in the Milky Way and LMC by scanning across
the clusters during exposures. Our training set also include spectra
of individual stars in those GCs from which abundances can be obtained
in the normal way to provide a check on our integrated--light results.
We present here the preliminary integrated---light analysis of one GC
in our training set, NGC~104 (47~Tuc), and outline some of the
techniques utilized and problems encountered in that analysis.
\end{abstract}


\section{Introduction}             

\begin{figure}[!t]
\begin{center}
\includegraphics[width=250pt]{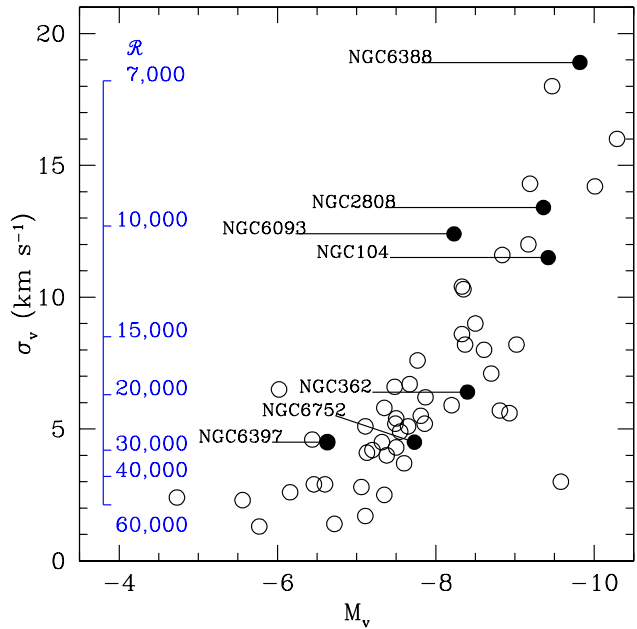}
\end{center}
\caption{ The distribution of Galactic GCs in absolute magnitude vs.
line--of--sight velocity dispersion (Pryor \& Meylan 1993). The inset
scale shows the limiting line--width (and thus spectral resolution,
${\sc R}=\lambda/\Delta\lambda$) obtainable for GCs as a function of
the line--of--sight velocity dispersion, $\sigma_v$. }
\label{fig:mvsigma}
\end{figure}

We are developing a method which will let us measure detailed chemical
abundances of Globular Clusters (GCs) from high--resolution, echelle
spectra of their integrated light. Our goal is to use the detailed
abundance patterns of GC systems in galaxies as distant as 4--5 Mpc to
constrain their formation and chemical enrichment histories 
in the same way as the abundance patterns of old stars have
been used to do so in the Milky Way (\textit{e.g.} 
Edvardsson \textit{et~al.}\ 1993;
McWilliam \textit{et~al.}\ 1995; 
Prochaska \textit{et~al.}\ 2003; 
Bensby \textit{et~al.}\ 2004ab; 
Fulbright, McWilliam \& Rich 2005).

The elements which are the most useful for understanding the processes
of galaxy evolution are those which are returned to the ISM primarily
through either
high--mass, core--collapse SNe (Type II = SNII) or through
low--mass, accretion--induced SNe (Type Ia = SNIa). The key is
that the former (SNII) evolve on time--scales of order 
$10^6$ years, while the latter (SNIa) generally take of order
$10^9$ years.  Elements produced by SNII ---
``$\alpha$--elements'' (\textit{e.g.} Si, S, Ca) and r--process elements
(\textit{e.g.} Eu) --- will therefore build up rapidly over the first few
megayears of star formation or in a starburst.  Elements produced
in SNIa and SNII --- Fe and Fe--peak elements (e.g. Sc, V, Cr, Mn, Fe,
Co, Ni) --- build up over many gigayears as the contribution from SNIa
increases.  The ratio [$\alpha$/Fe] relative to [Fe/H] is therefore a
rough constraint on the star formation rate.  Comparison with overall
abundance then constrains the timescale of formation.

Unfortunately, the Milky Way is currently the only galaxy for which
even a partial history of formation and enrichment can be traced in
the stellar fossil record through to the present day.  This is because
detailed abundance analysis requires high signal--to--noise ratios and
high spectral resolutions (S/N$\geq60$, $R>20,000$). Old stars are too
faint to be studied in this way beyond the nearest dwarf spheroidal
members of the Local Group ($\leq100$~kpc).  Even in the nearest
gas--rich irregulars of the Local Group, abundance measurements are
limited to a few very luminous supergiants (\textit{e.g.}\ Venn
\textit{et~al.}\ 2004), and results can only be obtained for a few
elements.  Moreover, these bright supergiants are very massive (young)
and so only identify the recent gas composition.  The history of a
galaxy can only be learned from long--lived, low--mass stars, which
are too faint for high--resolution spectroscopy beyond the nearest
dwarf spheroidal galaxies (\textit{e.g.}\ Shetrone \textit{et~al.}\
2003).

These limitations drive us to target Globular clusters (GCs).  Unlike
single stars, high--resolution spectra \emph{can} be obtained of
unresolved GCs out to $\sim4$~Mpc with \emph{current} telescopes.  GCs
are bright enough ($-10<M_v<-6$ mag) and have low enough velocity
dispersions ($2-20$ km/s) that even weak lines ($\sim15$ m\AA) could
be detected in spectra of their integrated light.  Moreover, several
lines of evidence from photometry and low--resolution spectroscopy
already suggest that GCs trace the star formation and the global
formation history of their parent galaxy: the number of GCs per galaxy
is roughly constant relative to the total light and mass of a galaxy;
young GCs are found in regions of active star formation
(\textit{e.g.}\ Schweizer \& Seitzer 1993, Barth
\textit{et~al.}\ 1995); and the properties of both metal--rich (red)
and now also metal--poor (blue) GC systems seem to correlate with the
properties of their parent galaxies (\textit{e.g.}\ Zepf \& Ashman
1993, Geisler, Lee \& Kim 1996; Gebhardt \& Kissler--Patig 1999; Kundu
\& Whitmore 2001; Larsen \textit{et~al.}\ 2001; Strader, Brodie, \&
Forbes 2004).  All of which implies that GCs are bright, observable
markers of the chemical enrichment record of normal galaxies.

Of course, significant progress has also been made in understanding
the chemical enrichment histories of elliptical galaxies and bulges
from analysis of low--resolution, integrated--light (IL) spectra using
line indexes, such as the Lick system (Burstein \textit{et~al.}\ 1984,
Faber \textit{et~al.}\ 1985, Worthey \textit{et~al.}\ 1994, Worthey \&
Ottaviani 1997, Trager \textit{et~al.}\ 1998).  These indexes provide
estimates of age and composite ``metallicity'' (``Z'') based on lines
from multiple elements such as Fe, Mg, and Ca (see Rose 1984, Worthey
\textit{et~al.}\ 1994, and Worthey \& Ottaviani 1997).  GCs are ideal
targets for these index systems as they are nearly ideal, single--age
stellar populations.

A great deal of progress has recently been made with line index
systems by using various kinds of principle component analysis to
obtain better leverage on Z and also to obtain an ``E'' parameter,
which includes all of the elements (O, Mg, C, etc) typically found to
be enhanced in elliptical and bulge (rapidly forming) populations
(Proctor \& Sansom 2002; Proctor, Forbes, Beasley 2004; Strader \&
Brodie 2004).  However, even for GCs, information from line indexes
are limited by two difficulties.  First, E and Z do not have the
interpretive power of detailed abundances because they include
elements which form in multiple sites and include elements which are
known to be affected by GC self--enrichment (see discussions in
Gratton, Sneden \& Carretta 2004). These indexes do not entirely
isolate the [$\alpha$/Fe] ratios which constrain formation timescales.
Second, line index results are very sensitive to the calibration of
the system, and it can be hard to obtain stellar libraries of the
appropriate age and enrichment or to know what calibration to use.
This complication has been widely discussed in the literature
(\textit{e.g.}\ Trippico \& Bell 1995; Trager \textit{et~al.}\ 2000ab;
Thomas, Maraston, \& Bender 2003; Proctor \textit{et~al.}\ 2004; 
Tantalo \& Chiosi 2004).  Adjustments to the calibration system have
been published recently based on stellar models which include a range
of enrichments. However, not all of these agree with each other and
the models themselves are not well tested against observed
populations.  This concern is particularly relevant given recent work
on M31, which shows that the Milky Way GC system may not be generally
representative of GCs in even spiral galaxies; M31 appears to contain
a disk GC system (Morrison \textit{et~al.}\ 2004) which is at least partly
composed of young GCs (0.1--0.8 Gyrs; Barmby \textit{et~al.}\ 2000,
Beasley \textit{et~al.}\ 2004) with different abundance patterns than
are seen in the Milky Way (Burstein \textit{et~al.}\ 2004).

Fortunately, it \emph{is} possible to obtain high--resolution spectra
of unresolved GC.  A typical GC has velocity dispersions of
$2<\sigma_v<20$ km/s (Pryor \& Meylan 1993, see Figure 1), so that
spectra of their integrated light can have line widths in the range
$60,000>\lambda/\Delta\lambda>6,500$.  For comparison, an elliptical
has $\sigma_v\sim 200$ km/s or greater and limiting line widths of
$\lambda/\Delta\lambda \sim 1500$.  Absorption features as weak as a
$\sim15$ m\AA\ can therefore be seen in high--resolution, {\it integrated
 light} spectra of all but the most massive GCs  with S/N$\approx
60-90$.

Detailed abundances have never been obtained for unresolved GCs 
because current methods of abundance analysis only work for individual
stars.  We describe here progress that we have made in adapting the
basic techniques of single star abundance analysis to spectra of
integrated--light.

\section{A Method for Light--Weighted Abundance Analysis}
\label{sec:method}

Abundances are not directly observable quantities; the strength of any
given absorption line is a function not only of the abundance of the
element, but also of the physical properties (e.g. mass and
temperature) of the star.  The method we are developing for analyzing
IL spectra is based on the standard techniques used to analyze
individual stars.

The standard technique for abundance analysis of single stars involves
comparing the observed equivalent widths (EWs) of lines for each
species with those predicted by model stellar atmospheres and spectral
line synthesis.  The model atmospheres have the following, observationally
constrained parameters: effective temperature ($T_{\rm eff}$, from
$B\!-\!V$), specific gravity ($\log g$, from luminosity), and
microturbulence ($\xi$).  In principle, the only free parameter is
abundance, [A/H].  So far, we have used Kurucz models in which Fe alone
is varied, so that [A/H] is roughly [Fe/H].  The Kurucz model
atmosphere grids (available from R. Kurucz at {\tt
http://cfaku5.harvard.edu}) give optical depth, temperature, pressure,
and electron density in each of 64 layers.  The EW of any particular
spectral line is then found by radiative transfer through these
layers. Given the model atmosphere, the EW of a line is uniquely
determined by the wavelength ($\lambda$), excitation potential (EP),
and {\it gf} value of the line, and by the abundance of the element in
question, $[X/H]$.  Of these, only [X/H] is adjusted.  In standard
analysis, the best model atmosphere is found by adjusting [Fe/H] in
the line synthesis and model atmospheres until a unique,
self--consistent solution is found from all Fe I and Fe II lines
($\sim 100$ lines).  A strong constraint on the correct model
atmosphere comes from requiring that Fe I lines over the full range of
$\lambda$, EP, and EW give a stable Fe solution.  This requirement is
used to adjust $T_{\rm eff}$, $\log g$, and $\xi$  to find the
appropriate stellar atmosphere model.  Lines from any other element
are then uniquely determined by the input abundance of that element
([X/H]) in the line synthesis step.  The right abundance is found by
adjusting this value until predicted EWs match those observed.

\begin{figure}[!ht]
\begin{center}
\includegraphics[width=250pt]{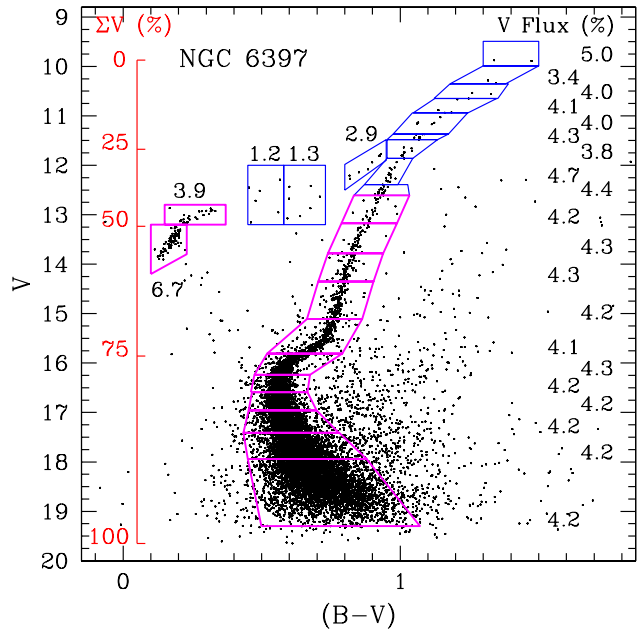}
\end{center}
\caption{A color--magnitude diagram for NGC~6397 (Kaluzny 1997).  The
scale on the left shows the cumulative $V$--band flux coming from
bright to faint stars.  The boxes sub--divide the stellar population
into groups containing $\sim5$\% of the $V$--band light, as indicated.
An analysis based on this CMD, for example, would
include one stellar atmosphere model per box.}
\label{fig:6397cmd}
\end{figure}

To interpret an IL spectrum, we have developed an original method of
detailed chemical abundance analysis which involves producing a
light--weighted EW, building on those techniques used for individual
stars. For a {\it resolved} Galactic GC, the observed CMD tells us the
exact fraction of light coming from every stellar type in the
cluster. For NGC~6397, for instance, Figure 2 shows the CMD of
NGC~6397 with boxes indicating groups of stars which each contain
roughly 5\% of the cluster's light.  Based on this observed CMD, we
can compute a model atmosphere for each box using the Kurucz model
atmospheres (ATLAS9) for a range in abundance, [A/H].  We have
developed code which then synthesizes the individual Fe lines by
iteratively calling the spectrum synthesis program MOOG (Sneden 1974)
and varying the Fe in the line synthesis only until consistent
solutions are obtained.  With this Fe abundance, we can then compute a
final stellar atmosphere for each box. We then identify the abundances
of other elements by adjusting [X/H] (in the line synthesis only) to
match the observed EWs for a species.  In this analysis, a
microturbulence scaling law is adopted ($\xi \propto \log gf$).

\begin{figure}[!t]
\begin{center}
\includegraphics[width=130pt]{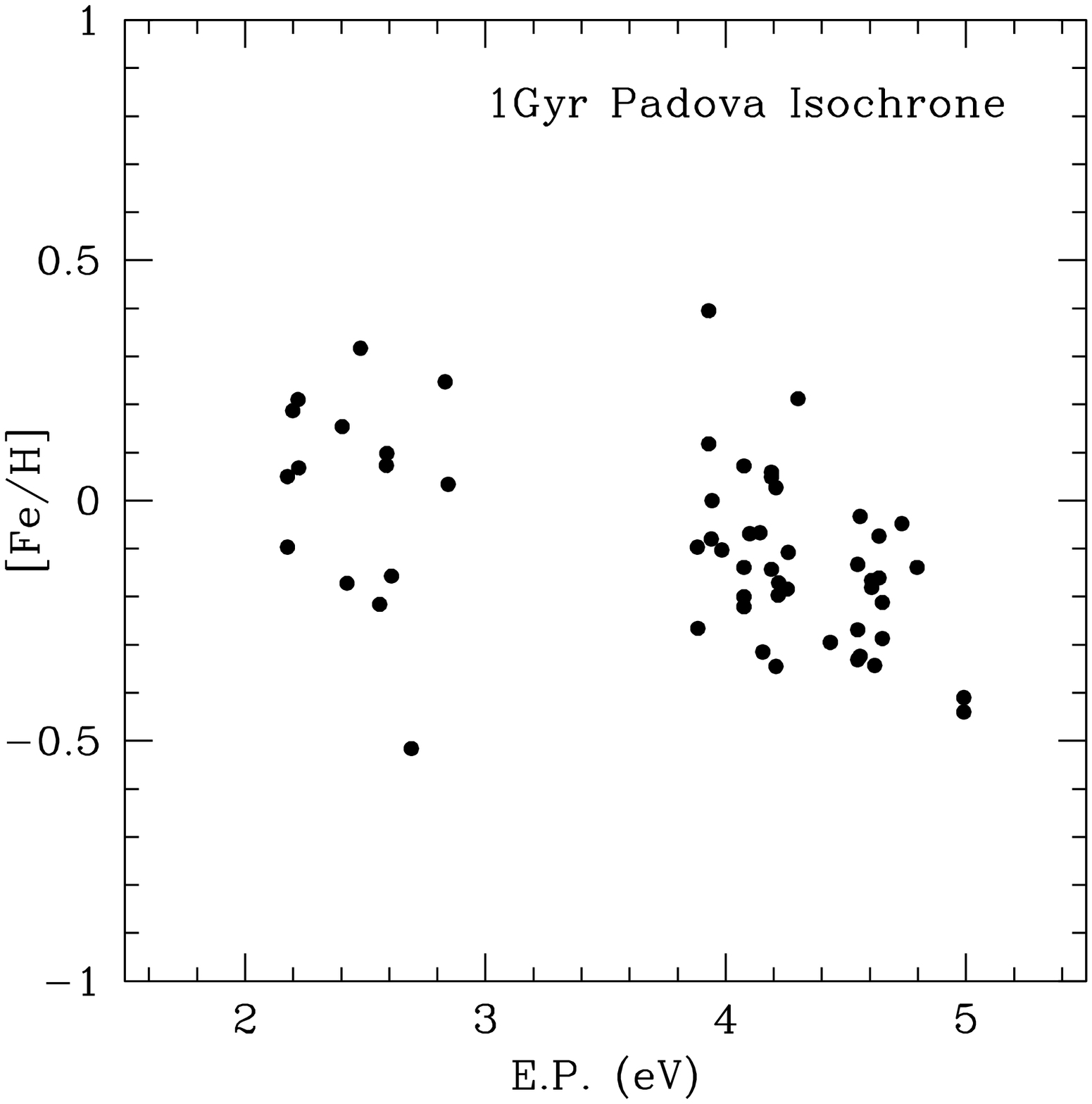}\hspace{-0.13in}
\includegraphics[width=130pt]{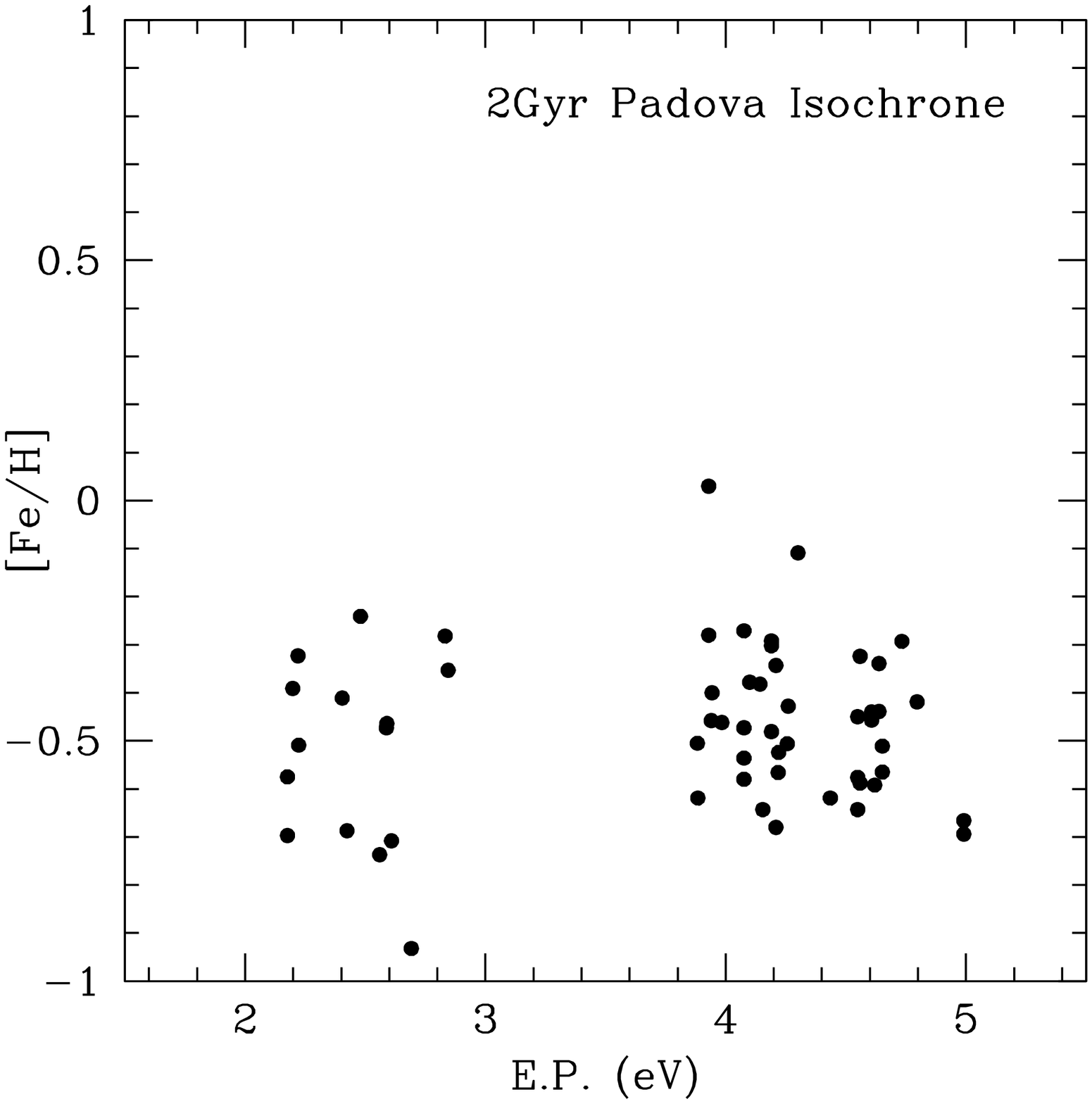}\hspace{-0.13in}
\includegraphics[width=130pt]{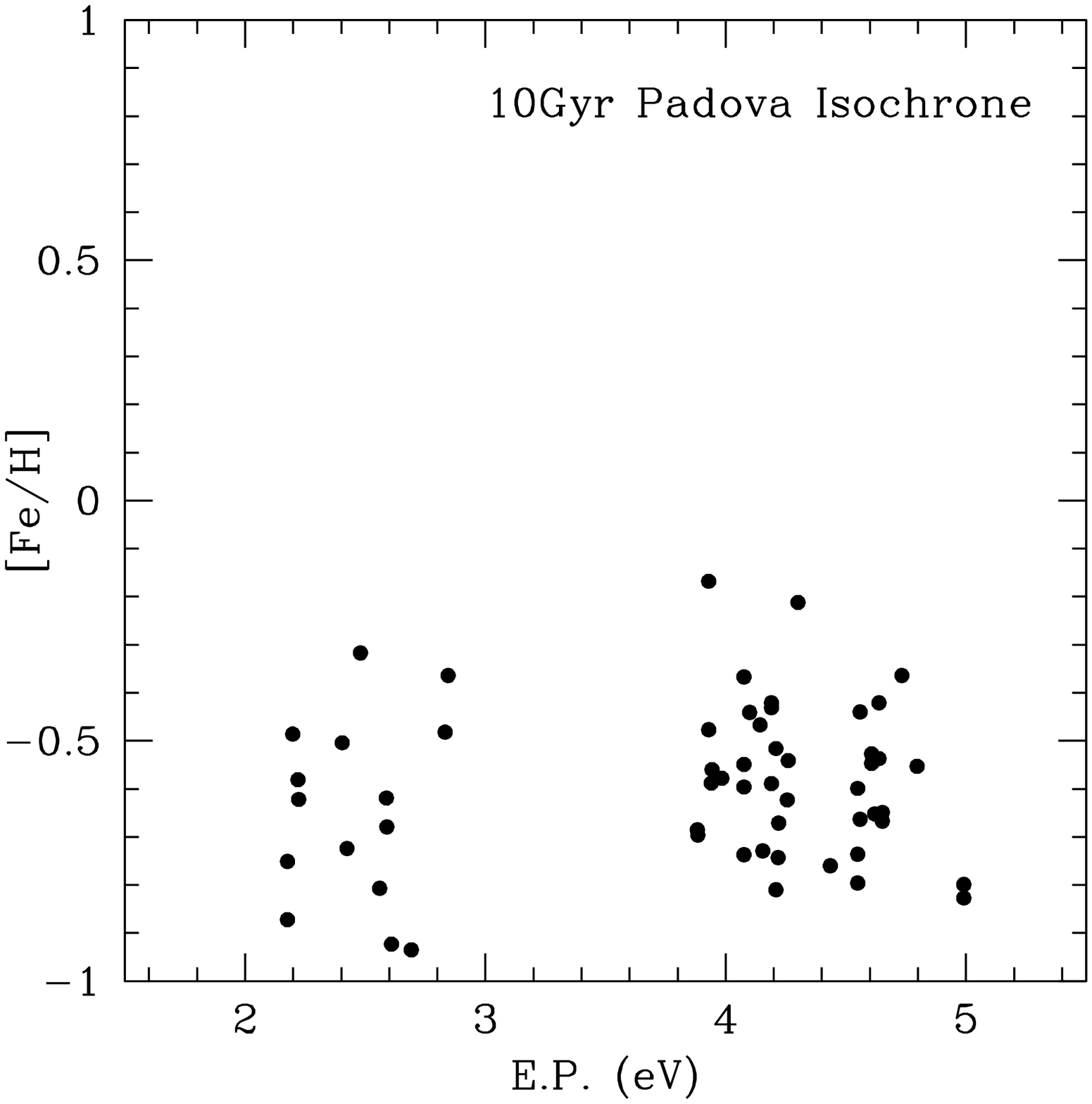}
\end{center}
\caption{Excitation potential vs [Fe/H] for all iron lines measured in
the IL of 47 Tuc.  Each plot shows the best--fit [Fe/H] solution for
one isochrone.  A systematic negative trend is clear in the 1 Gyr
isochrone, but weak in older isochrones.  Scatter is largely due to
uncertain \emph{gf} values.  It should be possible to obtain greater
precision by using \emph{gf} values tuned to the abundance solution of
Arcturus, a nearby metallicity reference star.  In RGB star analysis,
such techniques reduce the scatter in EP vs.\ [Fe/H] to $\pm0.05$ {\it
rms}.}
\label{fig:fevsep}
\end{figure}

As a first test of this basic strategy, we have analyzed the
integrated light spectrum of NGC~6397 using its observed CMD (see
Figure 2) to define the contributing stellar atmosphere models and
their weights.  From this analysis, we successfully obtain abundances
for Fe--peak, $\alpha$--elements, and light elements which are in good
agreement with the results from other groups for single stars in
NGC~6397 (see Bernstein \& McWilliam 2001 and Bernstein \& McWilliam
2005).  However, when we target extragalactic GCs, we will not be
able to use resolved CMDs to empirically identify the right mix of
stellar populations.  We must therefore learn to use theoretical CMDs
--- isochrones based on stellar evolution tracks --- to analyze the
IL spectra. To do this, and to understand the limitations and impact
of stellar evolution models themselves, we have observed a set of
GCs in the Milky Way and LMC to serve as a ``training set''.

\section{A Training Set of Galactic and LMC GCs}
\label{sec:tset}

Our ``training set'' consists of seven Milky Way and eight LMC
GCs. These span the range of velocity dispersions, abundances,
ages, and HB morphologies available in the Milky Way and LMC systems
(see Figure \ref{fig:mvsigma} and Table \ref{table:tset}).  For each
of the training set GCs, we have obtained integrated light spectra by
uniformly scanning the central $32\times32$ arcsec$^2$ and
$12\times12$ arcsec$^2$ in the Milky Way and LMC GCs, respectively.
Crucially, these clusters are all spatially resolved, so detailed
color magnitude diagrams (CMDs) can provide age estimates and \emph{a
priori} knowledge of the member stellar populations, including HB
morphology.  We can use this information to refine our methods and
determine how such variables affect our results.  Moreover, we can
also identify individual red giant branch (RGB) stars from these CMDs
and use spectra of these individual stars to obtain ``fiducial''
abundances by standard methods which are consistent with our own
atmospheric modeling and synthesis for the integrated light.  Below
we describe the results we have obtained for one Milky
Way GC: NGC~104 (47~Tuc).

To analyze 47 Tuc as we would an unresolved GC, we use isochrones as
the template for its stellar population.  The analysis here uses
isochrones based on the stellar evolution tracks of the Padova group
(Girardi \textit{et~al.}\ 2000) and a Kroupa IMF (Kroupa \& Boily
2002).  We complete the same analysis for isochrones with a range of
age and abundance (1--16 Gyrs, $-2.3<$[Fe/H]$<0.2$).  The first
question one might ask is whether the resulting $\sim15$ Gyr,
[Fe/H]$\sim -0.7$ isochrone \emph{looks} like the observed CMD of
47~Tuc.  Two differences are evident, although not unexpected
(\textit{e.g.}\ Bergbusch \& VandenBerg 2001): the isochrones include
lower mass stars than are observed in the clusters and do not
reproduce the observed AGB bump. The same problems exist with the
BASTI (Pietrinferni \textit{et~al.}\ 2004) stellar evolution models.
To begin, we use the Padova isochrones ``as is,'' with no adjustments
for mass segregation or the AGB bump.

\begin{figure}[!t]
\begin{center}
  \includegraphics[width=375pt]{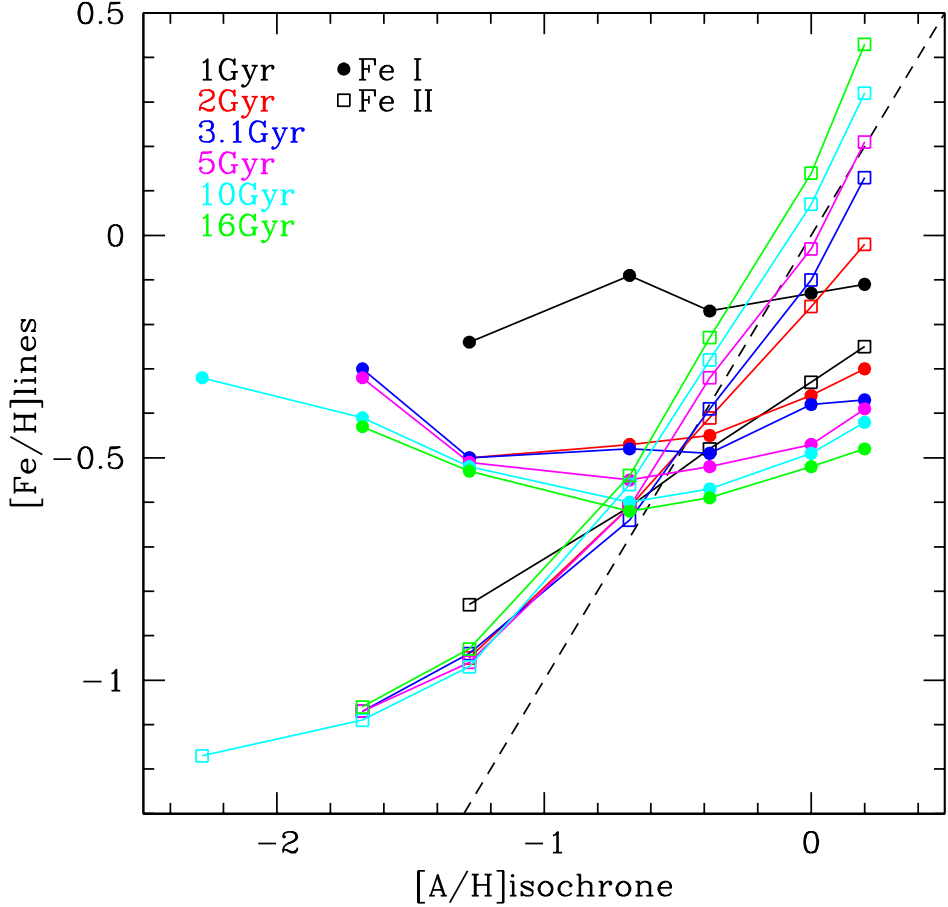}
\end{center}
\caption{ The inferred [Fe/H] abundance from Fe I and Fe II lines for
Padova isochrones of age 1--16 Gyrs and abundances of [A/H]$=-2.35$ to
0.2.  These isochrones do not include $\alpha$ enhancement.  The Fe I
lines give stable solutions with $\lambda$ and EP for isochrones with
ages older than about 3 Gyrs (see Figure \ref{fig:fevsep}).  The Fe II
and Fe I lines give the same solution (the lines connecting Fe I and
Fe II solutions cross) at a narrow range in abundances, namely at
[Fe/H] = $-0.63 \pm 0.05$.}
\label{fig:FeIFeII}
\end{figure}

For each isochrone, we split the stellar population into roughly 20
groups each containing roughly 5\% of the light, similar to the boxes
shown in Figure \ref{fig:6397cmd}.  We then produce a stellar
atmosphere for each group and synthesize a light--weighted EW for
about 70 Fe I and 10 Fe II lines by combining the synthesized
EWs from MOOG for each group, as described in \S\ref{sec:method}
Note that with the isochrones, the parameters of the atmosphere model
are completely dictated by the parameters of the isochrone.  We adjust
only the abundance of Fe in the line synthesis step to obtain a good
match between observed and predicted EWs for all Fe I and Fe II
line. Figure \ref{fig:fevsep} shows the best--match Fe value for
Fe I lines as a function of excitation potential for three
different isochrones.  The average of these solutions gives us the
inferred [Fe/H] value (and a statistical uncertainty) for each
isochrone.  Preferred solutions have small {\it rms} scatter and no slope
with excitation potential, wavelength, and EW.  We then do the same
test for Fe II lines.  A plot showing the inferred value of [Fe/H] from
Fe I and Fe II lines for each isochrone is shown in Figure
\ref{fig:FeIFeII}.  Surprisingly, we find that best--fit [Fe/H]
solution from Fe I lines is nearly independent of the input isochrone
{\it metallicity}, while Fe II lines are nearly independent of the
input {\it age}.  Together, the Fe I and Fe II lines provide an
excellent constraint on the abundance of the cluster and a way to
break the age--metallicity degeneracy that will never be achievable at
low resolution.

From the best--fit isochrone, we get [Fe/H]=$-0.63\pm0.03$ ($1\sigma$
statistical uncertainty).  We then adjust the luminosity function of
one isochrone to remove low--mass stars and add an AGB bump.  With
this adjusted isochrone, we find [Fe/H]$=-0.73$.  Note that these
results differ by only 0.1 dex, and that both results are roughly
within the range of values obtained recently by different groups from
individual stars (Kraft \& Ivans 2003, Carretta \textit{et~al.}\
2004).  Nevertheless, the change in our [Fe/H] solution with these
adjustments emphasizes the importance of the luminosity function to
our results.  We will use the rest of the training set to determine the
best approach to dealing with these differences between isochrones
and real CMDs.

We use the adjusted isochrone to then obtain abundances for a broad
range of key elements (see Table \ref{table:47tuc}).  Our final
abundances are in good agreement with those in the literature for
individual 47 Tuc stars.  Moreover, we find the expected abundances
patterns $\alpha$-- (enriched relative to solar), Fe--peak (roughly
solar), and r--process (enriched) elements that one would expect for
an old globular cluster.  We also detect enrichment of Na and Al,
which is a consistent with self--enrichment through proton--burning in
AGB stars and is found for some individual stars (with star--to--star
variations) in 47 Tuc (Carretta \textit{et~al.}\ 2004). This suggests
that we will be able to measure such abundance variations in
unresolved clusters even if present in only a fraction of the stars in
a given cluster!  Finally, we note that the age of 47 Tuc is only
constrained from our analysis to be greater than $\sim3$ Gyrs. We may
be able to improve this by calibrating the $\log gf$ values of
individual Fe lines to the abundance solution for Arcturus; doing
so should reduce the scatter in plots like those shown in Figure
\ref{fig:fevsep} to $\pm 0.05$, as it does in the analysis of
individual stars (see Bernstein \& McWilliam 2005).  However, it
is also clear from evolutionary tracks and observed CMDs of Milky Way
GCs that stellar populations themselves simply do not vary much with
age after a few Gyrs (\textit{e.g.}\ Yi \textit{et~al.}\ 2003).

\section{Further Spectroscopic Constraints on Integrated Light Abundances}

An important issue to explore is the influence of Horizontal Branch
(HB) morphology on our abundance analysis because it is not uniquely
correlated with age or metallicity and therefore not reliably
characterized by the isochrones.  Balmer line EWs \emph{and profiles}
will be helpful in this regard, because they are very sensitive to the
light fraction contributed by hot stars (\textit{e.g.}\ Bernstein \&
McWilliam 2002; Schiavon \textit{et~al.}\ 2004; Bernstein \& McWilliam
2005;).  To this end, we have compared observed Balmer profiles for
several of our training set clusters with synthesized, light--weighted
Balmer profiles based on the isochrones. Our preliminary work suggests
that we will need to also synthesize blended lines in the Balmer line
wings in order to get accurate Balmer EWs and profiles, particularly
in the metal rich clusters. This may be a particular problem for
low--resolution spectra. Similar issues have already been noted in the
literature to the extent that Balmer line indexes in low--resolution
spectra seem to be sensitive to the calibration of the line system
(\textit{e.g.}\ Proctor, Forbes \& Beasley 2004).  Additional
constraints may also come from CaII lines, which have also been shown
to track HB color in low--resolution spectra (e.g. Burstein
\textit{et~al.}\ 2004; Proctor, Forbes \& Beasley 2004).  Independent
models for horizontal branch morphology can also be combined with the
isochrone models directly to explore the impact on inferred abundances

We are also planning to explore a broader range of parameters in the
isochrones themselves as more models become publicly available.  For
example, stellar evolution tracks with different abundance ratios
(e.g. enriched in $\alpha$--elements or with different helium
fractions) may yield systematically different abundances and should
also be tested against the training set.

\section{Applications}

It is already possible to obtain spectra of extragalactic GCs that can
be analyzed with this technique.  In the local group, abundances of
the more distant LMC GCs could be obtained in a few hours with 4--m
class telescopes taking IL spectra, instead of requiring many hours of
exposure time on 8--m class telescopes observing individual cluster
stars.  Beyond the LMC, an handful of galaxies within 4--5~Mpc can be
observed using using current 6.5--10~m telescopes.  We have already
obtained a few spectra ($R\sim 20,000$, S/N$=60-900$ at $H\alpha$) of
confirmed GCs in galaxies NGC~1313 and NGC~5128 (both $\sim4$~Mpc
away) with the MIKE echelle spectrograph (Bernstein
\textit{et~al.}\ 2003) on the Magellan Telescopes.  In the future,
hundreds of galaxies in the local universe could be observed with the
next generation of ground--based 20--30~m telescopes.  In addition,
the abundances and integrated light spectra we already have in hand
can be used to better understand and calibrate the low--resolution
data and line--index systems.  This is important because line indexes
will always be able to reach galaxies at greater distances than
the high--resolution technique described here.

\acknowledgements We gratefully acknowledge the help of Las Campanas
technical staff and particularly telescope operators Fernando Peralta
and Herman Olivares. R.~A.~B.\ acknowledges partial support from NASA
through Hubble Fellowship grant HF--01088.01--97A awarded by Space
Telescope Science Institute, which is operated for NASA by the
Association of Universities for Research in Astronomy, Inc.\ under
contract NAS 5--2655.  A.~M.\ acknowledges support from NSF grants
AST--96--18623 and AST--00--98612.  We also thank the organizers for a
very enjoyable and stimulating conference.

\setcounter{table}{0}

\begin{landscape}
\begin{table*}[!h]
\renewcommand{\arraystretch}{1.0}
\begin{center}
\caption[]{\footnotesize 
`Training set'' clusters in the Milky Way and LMC.}
\begin{tabular}{l c c c c c c c c c c }
\scriptsize
\hfil \\
\hline
\hline
GC      & RA    &  Dec  & $\mu_V$  & $V_{\rm tot}$ & $M_{V{\rm tot}}$ & HBR$^a$   & $ R_\odot$ & ${r_{\rm core}}^b$ & $\sigma_v{\small\rm core}$  &  [Fe/H]\\
        &{(J2000)}&(J2000) & ${{\rm mag}/{\rm asec}^{2}}$ & mag & mag      & & kpc        & pc              & km/s        &      \\
\hline
\multicolumn {6}{l}{\underline{Milky Way clusters}} &  & \\ 
%  name      ra             dec            mu-v     v-tot   Mv    B-R/B+V+R   Rsun      rcore     sigma   fe/h
NGC 104	&  00 24 05.2  	&  --72 04 51  	&  14.42  & 3.95  & --9.42   & --0.99    &   4.5    & 0.40    &  9.8 & --0.65 \\
NGC 362	&  01 03 14.3  	&  --70 50 54 	&  14.79  & 6.40  & --8.40   & --0.87    &   8.5    & 0.17c   &  6.3 & --1.16 \\
NGC 2808&  09 12 02.6   & --64 51 47     &  15.17  & 6.20  & --9.39   & --0.49    &   9.6    & 0.26    & 14.2 & --1.15 \\
NGC 6093&  16 17 02.5	&  --22 58 30	&  15.38  & 7.33  & --8.23   & 0.93     &  10.0	  & 0.15    & 12.5 & --1.75 \\
NGC 6388&  17 36 17.0  	&  --44 44 06  	&  14.50  & 6.72  & --9.82   &          &  11.5    & 0.12    & 18.9 & --0.60 \\
NGC 6397&  17 40 41.3  	&  --53 40 25  	&  16.66  & 5.73  & --6.63   & 0.98     &   2.3    & 0.05c   &  3.3 & --1.95 \\
NGC 6752&  19 10 51.8 	& --59 58 55 	&  15.20  & 5.4   & --7.73   & 1.00     &   4.0    & 0.17c   &  4.5 & --1.83 \\
\hfil \\
\multicolumn {7}{l}{LMC clusters} & log(Age) & \\ 
\hline
\multicolumn{7}{l}{{\underline{Old ($>5$Gyrs)}}}  \\
NGC 1916 	& 5 18 39.00   & --69 24 24     & 15.8   & 9.88  & --8.96  & & $>$9	& $<1.5$& $8.2\pm1.2$ 	&--2.08   \\
NGC 2019 	& 5 31 56.7    & --60 09 33     & 15.5  	& 10.7 	& --7.94  & & $>$9	& 0.9	& $7.5 \pm 1.3$	&--1.3   \\
NGC 2005 	& 5 30 10.36   & --69 45 09     & 15.8 	& 11.57 & --7.48	 & & $>$9	& $1.3$ & $8.1 \pm 1.3$	&--1.92   \\
%\\
\multicolumn{9}{l} {{\underline{Intermediate age (0.1--1.5G yrs)}}} \\
NGC 1866 	&  5 13 39     & --65 27 54     & 18.5   & 9.76  & --8.74	 & & 8.1	&11.7   & ...		& --0.51 \\%\pm0.08^c$ \\ 
NGC 1978 	&  5 28 45.00  & --66 14 12     &        & 10.7	& --7.7	 & & 9.3	& ...   & $3.0\pm 0.5$ 	& $-0.91$\\%\pm0.15^c$  \\ 
%\\
\multicolumn{9}{l} {{\underline{Young ($\sim10$ Myr)}}} \\
NGC 1711  	& 4 50 37     & --69 59 06     & 17.0    & 10.11 & --8.3	 & & 7.4	& 5.7	&  ...	        & $-0.6$  \\ 
NGC 2002	& 5 30 21     & --66 53 00     & 16.0	& 10.1	& --8.3 	 & & 7.2 	&3.5	& ... 	        &  ...    \\ 
NGC 2100 	& 5 42 08     & --69 12 42     & 17.0    & 9.6   & --8.8   & & 7.2	&6.7	& ...	        & ...     \\ 
\hline
\hline
\label{table:tset}
\end{tabular}
\end{center}
\noindent{\footnotesize Notes: (a) HBR, the horizontal--branch ratio,
  is defined as $(B-R)/(B+V+R)$.  (b) A ``c'' indicates that the
  cluster is core--collapsed.  For the Milky Way GCs, values for 1--D
  velocity dispersion, $\sigma_v$, are from Pryor \& Meylan (1993);
  all other values are from Harris (1996).  Data for the LMC clusters
  comes from various sources, including Johnson \textit{et~al.}\ 2004, Hill
  (2004), Olsen \textit{et~al.}\ 1998, Olszewski \textit{et~al.}\ 1996
  and references therein.}
\end{table*}
\end{landscape}

\begin{table}[!ht]
\renewcommand{\arraystretch}{0.7}
\begin{center}
\caption[]{\footnotesize Element Abundances for 47 Tuc from 
Integrated Light Spectra}
\begin{tabular}{l c c c c c| c }
\scriptsize
\hfil\\
\tableline
\tableline
& & & & & &  \\  
{ Species, X} & 
{ $\epsilon(X)\,^{a}$} & 
{ $\sigma$} & 
{ $N_{\rm lines}$} & 
{ $(\sigma/\sqrt{N})^b$} & 
{ [$X$/Fe]$\,^c$ }& 
{ [$X$/Fe]$\,^d$ } \\
& & & & & &  \\
\tableline
& & & & & &  \\
\multicolumn{6}{l} {\underline{Fe--peak}}\\
Sc II	& 2.53	& ...  	& 1	& ...  & +0.13	& +0.13   \\
V  I	& 3.40	& 0.31 	& 5	& 0.16 & +0.11	& +0.05\\
Cr I 	& 4.87	& 0.14	& 3 	& 0.10 & --0.19	& +0.11\\
Mn I	& 4.49	& 0.30	& 4	& 0.17 & --0.31	& --0.29\\
Fe I	& 6.78 	& 0.24	& 71	& 0.03 & --0.73	& --0.67, --0.63$^f$\\
Fe II	& 6.81	& 0.15	& 8	& 0.06 & --0.70	& --0.56, --0.58$^f$\\
Ni I	& 5.47  & 0.18	& 12	& 0.05 & --0.05	& +0.06\\
\multicolumn{6}{l} {\underline{$\alpha$--elements}}\\
Si I	& 7.16 	& 0.21 	& 6     & 0.09 	& +0.33 & +0.30 \\
Ca I	& 5.81	& 0.24	& 12	& 0.07	& +0.19 & +0.20 \\
Ti I$^g$& 4.55	& 0.24 	& 13	& 0.07	& +0.34 & +0.26 \\
Ti II$^g$& 4.68 & 0.16	& 3 	& 0.11 	& +0.44 & +0.38 \\
& & & & & &  \\
\multicolumn{6}{l} {\underline{Light elements with possible intra--cluster variations}}\\
%Na I    & 5.97	& 0.19  & 3	& 0.10	& +0.38 & +0.23 \\
Mg I	& 7.02	& ... 	& 1	& ... 	& +0.17 & +0.40 \\
Al I	& 6.19  & 0.02  & 2	& 0.02	& +0.43 & NA    \\
& & & & & &  \\
\multicolumn{6}{l} {\underline{Neutron--capture elements  (s--, r--process)}}\\
Y II	& 1.34 	& ...  	& 1 	& ...	& --0.19  & +0.49$^e$\\ 
Y I:	& 1.29  & ...  	& 1 	& ...	& --0.24  &  NA\\
Zr I	& 1.80  & 0.21	& 2 	& 0.21  & --0.08	 & --0.22$^e$\\
Ba II	& 1.41  & 0.09	& 3	& 0.05  & --0.11  & NA	\\
La II	& 0.57  & 0.38	& 2	& 0.26  & +0.05  & NA\\
Nd II	& 0.80	& ...  	& 1	& ...	& +0.01  & NA\\
Eu II	& --0.12 & ...  	& 1 	& ...	& +0.04  & +0.36$^e$\\
\tableline
\tableline
\label{table:47tuc}
\end{tabular}
\end{center}
\noindent{\footnotesize{Notes: (a) The abundance of
element $X$ is defined as  $\epsilon(X) = \log_{10}[{\rm N(X)/N(H)}] +12$.
(b) Realistic uncertainties probably lie between the {\it rms} error, 
$\sigma$,  and the error in the mean, $\sigma/\sqrt{N}$.
(c) $[X/{\rm Fe}] = \epsilon(X/{\rm Fe}) -\epsilon(X/{\rm Fe})_\odot $. 
For Fe I  and Fe II this column indicates [Fe/H].
(d) From Carretta \textit{et~al.}\ (2004).
(e) From Brown \& Wallerstein (1989).
(f) From  Kraft \& Ivans (2003) using the Kurucz models; 
also, values of --0.70 and --0.65 were obtained from Fe I and Fe II, 
respectively, using the Gustafsson \textit{et~al.}\ (1975) models.
(g) Ti can also be considered an Fe--peak element and does 
show different abundance ratios than Si in Galactic GCs
(Lee \& Carney 2002).}}
\end{table}

\end{document}